\documentstyle[12pt,epsf]{article}
\def\beq{\begin{equation}}
\def\eeq{\end{equation}}
\def\bea{\begin{eqnarray}}
\def\eea{\end{eqnarray}}
\def\bef{\begin{figure}}
\def\enf{\end{figure}}
\def\S{{\bf S}}

\def\C{{\bf C}}
\def\Z{{\bf Z}}
\def\R{{\bf R}}
\def\P{{\bf P}}

\def\RP{{\bf RP}}
\def\CC{{\cal C}}

\def\CN{{\cal N}}
\def\CO{{\cal O}}
\def\CS{{\cal S}}

\def\Wt{{W_{\mbox{tree}}}}
\def\We{{W_{\mbox{eff}}}}
\def\Tr{{\mbox{Tr}}}
\def\eit{{e^{i\theta}}}
\def\ejt{{e^{-i\theta}}}

\def\ba{\begin{array}}
\def\ea{\end{array}}
\def\bce{\begin{center}}
\def\ece{\end{center}}

\def\ev#1{\langle#1\rangle}

\def\pa{\partial}

\def\IC{{\relax\hbox{$\inbar\kern-.3em{\rm C}$}}}
\def\ID{\relax{\rm I\kern-.18em D}}
\def\IE{\relax{\rm I\kern-.18em E}}
\def\IF{\relax{\rm I\kern-.18em F}}
\def\IG{\relax\hbox{$\inbar\kern-.3em{\rm G}$}}
\def\IGa{\relax\hbox{${\rm I}\kern-.18em\Gamma$}}
\def\IH{\relax{\rm I\kern-.18em H}}
\def\II{\relax{\rm I\kern-.18em I}}
\def\IK{\relax{\rm I\kern-.18em K}}

\def\IQ{\relax\hbox{$\inbar\kern-.3em{\rm Q}$}}

\begin{document}
\begin{titlepage}
\rightline{HUTP-01/A026} \rightline{HU-EP-01/22} \rightline{hep-th/0106040}
\def\today{\ifcase\month\or
January\or February\or March\or April\or May\or June\or July\or August\or September\or October\or November\or
December\fi, \number\year} \vskip 2cm \centerline{\Large\bf Open/Closed String Dualities and Seiberg Duality}
 \vskip 1mm
\centerline {\Large\bf  from Geometric Transitions in M-theory} \vskip 1cm \centerline{\sc Keshav Dasgupta
$^{a,}$\footnote{keshav@ias.edu}, Kyungho Oh$^{b,}$\footnote{On leave from Dept. of Mathematics, University of
Missouri-St. Louis, oh@hamilton.harvard.edu} and Radu Tatar$^{c,}$\footnote{tatar@physik.hu-berlin.de}} \vskip
1cm \centerline{{ \it $^a$ School of Natural Sciences, Institute for Advanced Study, Princeton NJ 08540, USA}}
\vskip 1mm \centerline{{ \it $^b$ Lyman Laboratory of Physics, Harvard University, Cambridge, MA 02138, USA}}
\vskip 1mm \centerline{{\it $^c$ Institut fur Physik, Humboldt University, Berlin, 10115, Germany}} \vskip 2cm
\centerline{\sc Abstract} \vskip 0.2in We propose a general method to study open/closed string dualities from
transitions in M theory  which is valid for a large class of geometrical configurations. By T-duality we can
transform geometrically engineered configurations into ${\cal N} = 1$ brane configurations and study the
transitions of the corresponding branes by lifting the configurations to M-theory. We describe the transformed
degenerated M5 branes and extract the field theory information on gluino condensation by  factorization of the
Seiberg-Witten curve. We also include massive flavors and orientifolds and discuss Seiberg duality which appears
in this case as a birational flop. After the transition, the Seiberg duality becomes an abelian electric-magnetic
duality. \vskip .2in \leftline{June 2001}

\end{titlepage}
\newpage

\section{Introduction}
Beginning with Maldacena's AdS/CFT conjecture \cite{mal}, the duality between gauge theories and gravity has been
actively studied by using the open/closed string theory dualities, mainly on non-compact singular or compact
Calabi-Yau $n$-folds. Geometric engineering and `brane engineering' are two of the main useful techniques.

Geometric engineering \cite{veng1,veng2} is a very powerful tool, but sometimes it is hard to manipulate due to
rigid holomorphic structures. On the other hand, brane  configurations, typically in the flat geometric
background, are relatively easy to manipulate and, by utilizing Witten's MQCD methods \cite{wit,witqcd}, it has
been very successfully used in studying the non-perturbative dynamics of low energy supersymmetric gauge
theories.

Often T duality has been applied to go from the  geometric set-up to the brane configuration set-up (with NS
branes and D branes) \cite{kls} but the results were obtained for semi-localized configurations where the NS
branes were smeared on some compact directions. In this paper, we systematically utilize T duality and Witten's
MQCD \cite{wit,witqcd} to investigate the $\CN =1$ large N duality proposed by Vafa \cite{vafa,civ}. Because we
consider the T-dualities given by a $\C^*$ action on smooth Calabi-Yau spaces, our solutions are fully localized,
although we do not write explicit supergravity solutions. The NS branes are fully localized because their
directions are identified with some fixed lines in the geometry, as we will see in the detailed discussion of the
T-duality. The smooth Calabi-Yau spaces considered here are either versal deformations or K\"ahler deformations
of singular Calabi-Yau spaces with only conifold singularities.

Recently, using previous results on  Chern-Simons/topological strings duality \cite{gova}, Vafa suggested a dual
picture in the large N limit between open string theory on D-branes wrapped over the $\P^1$ cycle of a resolved
conifold and a closed string theory on the deformed conifold where the D-branes have disappeared and have been
replaced by RR fluxes through the $\S^3$ cycle, together with NS fluxes through the dual noncompact 3-cycle
\cite{vafa}. The conifolds have been studied extensively in the last years in view of AdS/CFT duality and they
play a role in understanding dynamics of ${\cal N} = 1$ supersymmetric theories \cite{kw}.

The solution has been generalized to more complicated non-compact geometries in \cite{sv,civ,eot}. The ten
dimensional transition appears as a geometric flop in M theory on a $G_2$ holonomy manifold \cite{amv,achar,en}.
The question is whether the transition in more complicated geometries as the ones of \cite{sv,civ,eot} (and some
other geometries build in the same way) can be lifted to flops in other  $G_2$ holonomy manifolds. The known
non-compact examples are sparce \cite{gibbons,gomis} \footnote{See \cite{cvetic,gubser,witati} for
generalizations of Ricci-flat matric on $G_2$ holonomy. For compact   $G_2$ holonomy manifolds used in recent
physics literature see \cite{Kachru,kaste}.}. It would be nice to find an alternative method to study the
transition from D-branes wrapped on 2-cycles of some geometry to fluxes through 3-cycles of some other
geometrical set-up.

Such an alternative method was proposed by us in \cite{dot} where we  considered Vafa's transition by using the
MQCD brane configurations \cite{wit,witqcd}. We started with $U(N)$ gauge group given by $N$ D5 branes wrapped on
the $\P^1$ cycle of a resolved conifold. Under a T-duality on the circle action of the resolved conifold gives a
brane configuration with $N$ D4 branes between two orthogonal NS5 branes. By lifting to M theory, this becomes a
single M5 brane which includes the Seiberg-Witten curve for $SU(N)$ theory, the $U(1)$ part of $U(N)$ being, in
general, decoupled. In our discussion, a crucial point was that for $N$ D5 branes wrapped on $\P^1$ cycle of a
resolved conifold, in the limit when the cycle is of very small size, there is a transition from the $U(N)$
theories on the branes to $U(1)$ theory on the bulk. In this limit, the Seiberg-Witten curve degenerates and it
describes a $U(1)$ theory.
The component $\Sigma$ of the M5 brane becomes a
``plane M5 brane". The coupling
constant of the $U(1)$ gauge group has a RG flow which can be shown to arise naturally from the cut-off applied
to regulate a divergent integral over the period of the degenerated $\Sigma$. This cut-off in the integral is
related to the UV cutoff of the dual gauge theory. Our results offer a check of the new point of view stated in
\cite{av1} concerning the fact that that the Chern-Simons/topological strings duality of \cite{gova} is more
natural for the $U(N)$ gauge group.

In \cite{dot}, we provided a rederivation of results already obtained by using $G_2$ holonomy manifolds. One can
ask whether the same method can be applied for more complicated geometries, where a  $G_2$ holonomy manifold is
not known. The answer is yes and this is the subject of our current paper. Our claim is that any geometry which
resembles the conifold can be treated in the same way as the conifold itself. These geometries involve D5 branes
wrapped on different $\P^1$ cycles which, after T-duality, give more complex brane configurations with orthogonal
NS branes and D4 branes between them and which then
can be lifted to a single M5 brane.

In the present paper we consider two such geometries. The first one is the geometry used in  \cite{civ} for an
${\cal N} = 2$ theory deformed by a superpotential with terms with different powers of the adjoint field and the
second one is a geometry describing $U(N)$ group with fundamental matter. For the second geometry, we discuss the
electric/magnetic Seiberg duality between $SU(N)$ and $SU(F-N)$ theories
($F$ being the number of flavors)
and provide some hints for the product
gauge groups. In the geometrical picture, the Seiberg duality appears as a flop transition, a fact which was
suggested in \cite{deoz,civ}, and here we provide a proof of this.

In discussing the brane configurations which correspond to the geometry of \cite{civ}, we study the transition
between M5 branes corresponding to the product of $SU(N_i)$ gauge groups and degenerate M5 branes
corresponding to a product of $U(1)$ groups. We make a clear identification of the parameters of the M5 branes in
terms of ${\cal N} = 1$ field theoretical scales. These parameters are related to different fluxes in the
geometrical side of the type IIB  transition. The degenerate M5 brane describes the Seiberg-Witten reduced curve
and its exact form is derived. Our result should be compared to the one in \cite{civ}, where the match between
the Seiberg-Witten reduced curve and the curve given by the period matrix of the geometry was checked only at the
highest order.

We also discuss the case when we have orientifolds and the results of the M5 brane transition is to be compared
with results of \cite{sv,eot}.

\section{Vafa's ${\cal N}=1$ Duality and Brane Configurations}
\subsection{Large $N$ Duality Proposal}
It has been proposed that there
exists a open-closed string duality between field theories on D branes wrapped on
$\P^1$ cycles and closed string theory on complex deformed Calabi-Yau manifolds \cite{vafa}. In \cite{civ} it has
been discussed that an ${\cal N} = 1$, $U(N)$ theory with adjoint and superpotential \bea\label{vafasuperpot}
W_{tree}= \sum_{p=1}^{n+1}~ {g_p\over p} \Tr \Phi^p \eea (obtained on D5 branes wrapped on several 2-cycles of a
resolved geometry) is dual to an $U(1)^n$ theory with superpotential \bea \label{hecchu} -{1\over 2\pi i}W_{eff}=
\sum_{i=1}^n ~ (N_i \int_{B_i}^{\Lambda_0}~\Omega + \alpha_i\int_{A_i}~\Omega) \eea In the formulae
(\ref{vafasuperpot}) and
(\ref{hecchu}), the variables $n$ and $g_p$ are given in terms of geometry of the Calabi-Yau, the quantities
$N_i$ are the total $H_R$ fluxes through the $A_i$ cycles and $\alpha_i$ are $\tau H_{NSNS}$ fluxes through the
$B_i$ cycles. This theory has a UV cutoff given by $\Lambda_{\CN=1} \equiv
\Lambda_0$
which we shall refer simply as $\Lambda$ henceforth.

Before going further we would like to make some comments on the existence of the $H_{NSNS}$ field. If we just
consider the proposal that D5 branes wrapped on various
 $\P^1$ cycles disappear and are replaced by the
supergravity background they create, then this would imply the existence of $H_{RR}$ only as  the $H_{NSNS}$ field
could be switched to zero. But this is {\it not}
the case and we can give two arguments for this:

(1) Vafa's original idea \cite{vafa} in type IIA picture was based on a geometry whose complex structure was not
integrable which implies the existence of a 4-form through the noncompact 4-cycle of a resolved conifold. This is
the NS 4-form which goes into the mirror dual into an NS 3-form through the noncompact 3-cycle of a deformed
conifold. Therefore, the NS 3-form has geometrical origin and cannot be switched off.

(2) In order to identify the geometrical cut-off  with the scale of the gauge theory, we need the term
$\frac{1}{g_0^2} S$ in the superpotential (where $g_0$ is the bare coupling constant) which comes from
integrating the $H_{NSNS}$ over the noncompact 3-cycle. So $H_{NSNS}$ cannot be turned off.

\subsection{Brane Construction and Geometric Transition}

In the previous section we studied the duality conjecture from the point of view of special geometry. There is an
alternative way to motivate this duality. And this is by using brane configurations. The conifold geometry can
also be studied from intersecting brane constructions by making a T-duality and going to type IIA. Let us discuss
this construction in some details since similar constructions will be used throughout the rest of the paper.

Consider an action $S_c$ on the conifold given by $xy-uv=0$ as:
 \bea \label{sc}
S_c:~~(e^{i\theta}, x) \to x ,~~ (e^{i\theta}, y) \to y~~(e^{i\theta}, u) \to e^{i\theta} u ,~~ (e^{i\theta}, v)
\to e^{-i\theta} v,~~ \eea The orbits of the action $S_c$ degenerates along the union of two intersecting complex
lines $y=u=v=0$ and $x=u=v=0$ on the conifold. This action can be lifted to the resolved conifold and deformed
conifold. As discussed in \cite{dot}, the T-dual picture for the D5 branes on the finite 2-cycle of the resolved
conifold will be a brane configuration with D4 branes along the interval with two NS branes in the `orthogonal'
direction at the ends of the the interval, the length of the interval being the same as the size of the rigid
$\P^1$ . For the deformed conifold, by taking the T-duality, we obtain an NS brane along the curve $u=v=0$ with
non-compact direction in the Minkowski space which is given by \bea \label{dns} xy = \mu \eea in the x-y plane.

The above is thus the brane constructions for a conifold, resolved conifold and deformed conifold. Now to study
Vafa's duality we appeal to Witten's MQCD construction \cite{wit,witqcd}. As we saw above, $N$ D5 branes wrapping
a 2-cycle of a resolved conifold map to $N$ D4 branes between two orthogonal NS5 branes separated along $x^7$
directions. We denote the directions of the two NS5 branes along $x^{0,1,2,3,4,5}$ and $x^{0,1,2,3,8,9}$
respectively. The D4 branes are along $x^{0,1,2,3,7}$.

When we lift this configuration to M-theory all the branes in the picture become a {\it single} M5 brane with
complicated world-volume structure. We define the complex coordinates: \bea \label{complexcoord} x = x^4+ix^5,
~~~y=x^8+ix^9,~~~ t = exp(-R^{-1}x^7+ix^{10}) \eea where $R$ is the radius of the $11th$ direction, the world
volume of the M5 corresponding to the resolved conifold is given by $R^{1,3} \times \Sigma$ and $\Sigma$ is a
complex curve defined,
 up to an undetermined constant $\zeta$, by
\bea \label{m5res} y = \zeta x^{-1}, ~~t = x^{N}\eea
 Now if we consider the limit where the size of $\P^1$ goes to
zero, then the value of $t$ on $\Sigma$ must be constant because
 $\Sigma$ is holomorphic and there is no non-constant holomorphic map
into $\S^1$. Therefore the  M5 curve make a transition
 from a ``space'' curve into a ``plane'' curve. From (\ref{m5res}), we
obtain $N$ possible plane curves \bea \label{m5plane} \Sigma_k:~~~t=t_0, ~~xy = \zeta \exp {2\pi i k/N}, \quad k
=0, 1, \ldots , N-1. \eea

This is the way we see Vafa's duality transformation. After the transition the degenerate M5 branes are no longer
considered as the M-theory lift of D4 branes. This is now a closed string background, and in fact if one looks at
the T-dual picture of the deformed conifold (\ref{dns}) then this is exactly the M-theory lift of the deformed
conifold.

\section{Large N Duality Proposal for $\CN =1$ Theory with adjoint and superpotential}
\subsection{Geometric engineering and Brane configuration}
In \cite{civ}, the large N duality conjecture of Vafa~\cite{vafa} has been generalized  to $\CN=1$ $SU(N)$ theory
with adjoint chiral superfield $\Phi$ and with tree-level superpotential \bea \label{wt} \Wt = \sum_{k=1}^{n+1}
\frac{g_k}{k} \Tr \Phi^k \eea The classical theory with superpotential (\ref{wt}) has many vacua, where the
eigenvalues of $\Phi$ are various roots of \bea W'(x) = \sum_{k=1}^{n+1} \frac{g_k}{k} x^k \equiv
g_{n+1}\prod_{k=1}^n (x-a_k). \eea In \cite{dot}, the case for $n=1$ in $\Wt$ has been considered.

Without the superpotential (\ref{wt}), the theory is just four dimensional $\CN =2$ Yang-Mills theory. This can
be geometrically engineered on the total space of the normal bundle $\CO(-2) + \CO(0)$ over $\P^1$. The $\CN =2$
$U(N)$ gauge theory is obtained on the world-volume of $N$ D5 branes wrapping the $\P^1$ \cite{civ}. The adjoint
scalar $\Phi$ is identified with the deformations of the brane in the $\CO(0)$ direction.

\begin{figure}
\centerline{\epsfxsize=80mm\epsfbox{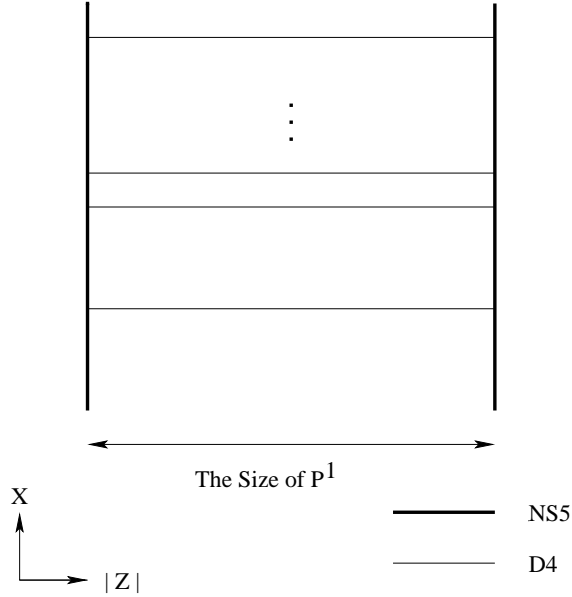}} \vspace*{1cm} \caption{The T-dual configuration of $N$ D5 on
$\P^1$ in $\CO(-2)+\CO(0)$ over $\P^1$.} \label{one}
\end{figure}

To describe the geometry in detail, we introduce two $\C^3$ whose coordinates are $Z, X, Y$ (resp. $Z', X', Y'$)
for the first (resp. second) $\C^3$. Then the total space of the normal bundle $\CO(-2) + \CO(0)$ over $\P^1$ is
given by gluing two $\C^3$'s with the identification: \bea \label{identif} Z' =1/Z, ~~ X' =X, ~~Y' =YZ^2 \eea Now
we will introduce a circle action and take a T dual along the orbits of the circle action. Consider a circle
action $S_2$: \bea (e^{i\theta}, Z) = e^{i\theta}Z, ~~(e^{i\theta}, X)
= X,~~(e^{i\theta}, Y) = e^{-i\theta}Y\\
\nonumber \label{s2} (e^{i\theta}, Z') = e^{-i\theta}Z', ~~(e^{i\theta}, X') = X',~~(e^{i\theta}, Y') =
e^{i\theta} Y'\eea Note that the action $S_2$ is compatible with (\ref{identif}) so that it is defined on
$\CO(-2) + \CO(0)$. Since the orbits of the action degenerate along $Z=Y=0$ and $Z'=Y'=0$, we have two NS branes
along $X$ direction at $Z=Y=0$ and $Z'=Y'=0$ after T-duality. Now if we take the T-dual of $N$ D5 wrapping
$\P^1$, it will be a brane configuration of $N$ D4 branes between two NS branes spaced along $X$ (Figure
\ref{one}). Two NS branes are parallel because the $X$  is coordinate of the trivial bundle $\CO(0)$ over $\P^1$.
\\
\\
The D4 branes can freely move along the direction of the NS branes corresponding to the direction in the Coulomb
branch, and the position of D4 brane is identified with the eigenvalues of the adjoint field $\Phi$. This is $\CN
=2$ $U(N)$ theory.

By adding the superpotential $\Wt$ (\ref{wt}), the theory will be broken to $\CN =1$ in which the NS brane is
curved into the $y$-space from  the straight line configuration in the $\CN =2$ theory.  We will only consider
the case where all vacua are non-degenerate i.e. $W'(X)$ has distinct roots. The T-dual of this $\CN =1$ brane
configuration can be geometrically engineered on a Calabi-Yau space. We consider a small resolution of a
Calabi-Yau with conifold singularities:
 \bea \label{cy} \CC:~~y(y-W'(x)) - uv=0 \eea
where $W'(x) = g_n\prod_{k=1}^{n}(x- a_k)$ with distinct $a_k$. Note that the singular points  are located at
$y=W'(x) =0$.  It will be shown that the singularities can be resolved by successive blow-ups which replace each
singular point by $\P^1$. The resolved space $\tilde {\CC} $ can be covered by two copies of $\C^3$, $U_p$,
$p=0,1$ with coordinates $Z, X,Y$ (resp. $Z', X', Y'$) which are related by \bea \label{gluedata} Z'  = Z^{-1},
\,\, X'=X,\,\, Y' = YZ^2 - W'(X)Z. \eea The resolution map $\sigma$ from the resolved space $\tilde {\CC}$ to the
singular Calabi-Yau space $\CC$ is given by
 \bea
x = X =X', \,\, y = YZ = Y'Z' +W'(X'), \,\, \\
u = Y = Z'(Y'Z' + W'(X')),\,\, v = Z(YZ -W'(X)) = Y'
 \eea

\begin{figure}
\centerline{\epsfxsize=60mm\epsfbox{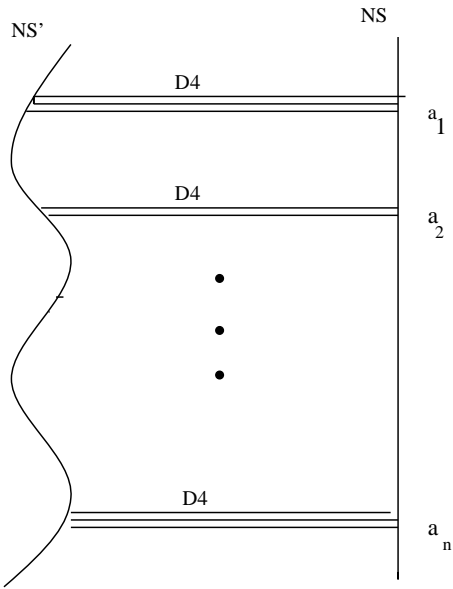}} \caption{The T-dual configuration of $N$ D5 distributed among $n$
$\P^1$.} \label{YYY}
\end{figure}

\begin{figure}
\centerline{\epsfxsize=60mm\epsfbox{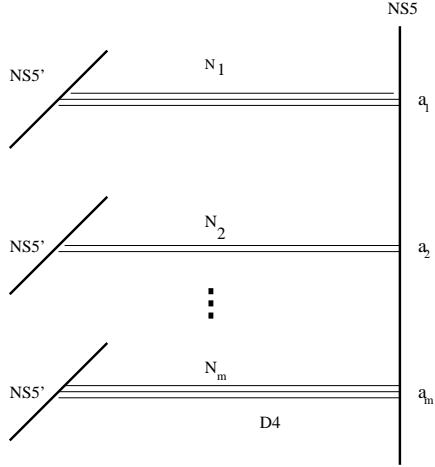}} \vspace*{1cm} \caption{The brane configuration in the limit
$g_n \to \infty$.} \label{XXX}
\end{figure}

The geometry change from the $\CN =2 $ theory (\ref{identif}) to the $\CN=1 $ theory (\ref{gluedata}) shows that
the $\P^1$ cycles in $\CN=1$ theory are fixed at $x =a_k$, $k=1, \ldots, n$ (which are the roots of $W'(x) =0 $)
while there is a family of $\P^1$ cycles in the $\CN=2$ theory. Since the circle action $S_2$ (\ref{s2}) is
compatible with the identification (\ref{identif}) for the $\CN=1$ geometry, the T-duality can be taken along
the direction of the orbits of the action $S_2$ on $\tilde{\CC}$. The orbits degenerate along $Z = Y =0$ and $Z'
= Y' =0$. Thus after T-duality, we have two NS branes which, under the resolution  map $\sigma$, map to $y=u=v
=0$ and $y = W'(x), u =v =0$ on $\CC$. Hence we obtain two NS branes, one straight NS brane denoted by NS and one
curved NS brane denoted by NS$'$ and the supersymmetry is broken to $\CN =1$. On the other hand,  D5 branes along
$\P^1$ cycles become D4 branes after T-duality.  If we distribute the $N$ D5 branes among the vacua $a_k$ by
wrapping $N_k$ branes on the $\P^1$ at $x=a_k$, this breaks the gauge group: \bea U(N) \to U(N_1) \times \cdots
\times U(N_n).\eea We obtain a brane configuration with two non-parallel NS branes  and D4 branes between them as
shown in (Figure \ref{YYY}) after T-duality. In the limit $g_{n} \to \infty$, the curved NS brane will be broken
into $n$ NS branes (Figure \ref{XXX}).
\\
\\
In \cite{civ}, the large N duality has been proposed via the geometric transition from the small resolution
$\tilde{\CC}$ of $\CC$ to a complex deformation of $\CC$. As we have $n$ isolated conifold singular points, the
complex deformation space (i.e. versal deformation space) is $n$ dimensional whose parameters are given as
$\mu_1, \ldots , \mu_n$. This can be obtained by adding a polynomial $f_{n-1}(x,\mu_1, ...\mu_n)$ of degree $n-1$
in $x$ with $f_{n-1}(a_k, \mu_1, \ldots , \mu_n) = \mu_k$ to (\ref{cy}), which is given by \bea \label{fn-1}
f_{n-1}(x,\mu_1, ...\mu_n)= \sum_{k=1}^{n} \mu_k\ \prod_{l=1,l\neq k}^n \frac{ x-a_l}{a_k-a_l}.\eea Hence the
deformed generalized conifold is given by \bea \label{cydef} \CC_s:~~ W'(x)y - f_{n-1}(x, \mu_1, \ldots , \mu_n)
-  uv =0.\eea The equation (\ref{cydef}) defines a variety in $\C^n \times \C^4$ where the coordinates of $\C^n$
are $\mu_1, \ldots , \mu_n$ and the smoothing $\Delta$ is given by the map induced by the natural projection:
\bea \Delta: \CC_s \to \C^n\eea and the fiber over the origin is the original singular Calabi-Yau space $\CC$
(\ref{cy}). On the generic fiber $\Delta^{-1}(\mu_1, \ldots , \mu_n)$, the $n$ three sphere $\S^3$ of size
$\mu_k$ will smooth out the singular point $x=a_k, y=u=v=0$. We consider a circle action $\CS_d$ on (\ref{cydef})
by \bea \label{csd} (\eit, x) \to x,~~(\eit, y)\to y, ~~(\eit u) \to \eit u,~~(\eit v) \to \ejt v. \eea Now if we
take T dual along the orbits of the action $\CS_d$, then NS brane will appear along $u=v=0$ which is given by
\bea \label{nsglobal} W'(x)y=f_{n-1}(x,\mu_1, \ldots , \mu_n ). \eea

\subsection{M theory interpretation}
In MQCD \cite{witqcd}, the classical type IIA brane configuration turns into a single fivebrane whose
world-volume is a product of the Minkowski space $\R^{1,3}$ and a complex curve in a flat Calabi-Yau manifold
\bea \label{M}M = \C^2 \times \C^*. \eea We denote the finite direction of D4 branes by $x^7$  and the angular
coordinate of the circle $\S^1$ in the 11-th dimension
 by $x^{10}$. Thus the NS branes are separated along the $x^7$
direction. We combine them into a complex coordinate \bea t = \exp ( -R^{-1}x^7 - i x^{10})\eea where $R$ is the
radius of the circle $\S^1$ in the 11-th direction.

Let us review the construction of the complex curve $\Sigma$ in the case of two orthogonal NS branes with $N$ D4
branes between them. On $\Sigma$ the coordinates of NS branes goes to the infinity only at their locations. So we
may identify $\Sigma$ with a punctured complex plane with coordinate $x$, and since there are $N$ D4 branes in
the type IIA picture, we should have $t = x^N$. Hence the $\Sigma \subset M$ is  an embedding of the punctured
$x$ plane $\C^{*}$ into the Calabi-Yau space $M$ by the map \bea \label{m5map} \C^{*} \longrightarrow \Sigma
\subset M, ~~~~~x \to (x, \zeta x^{-1}, x^{N}) \eea

The brane configuration of the generalized conifold consists of $(n+1)$ NS branes with $N_k$ D4 branes  between
the first and $(k+1)$-th NS branes. Hence the $\Sigma$ in this case will be an embedding of $n$-punctured complex
plane, coordinatized by $x$, into the Calabi-Yau space $M$. The $x$ coordinates of the punctures are given by
$a_k$ which is the location of the NS branes. Since there are $N_k$ D4 branes between the first and $(k+1)$-th NS
branes, we should have $t= (x-a_k)^{N_k}$ around $x =a_k$. Thus the $\Sigma \subset M$ is  an embedding of the
$n$-punctured $x$ plane  into the Calabi-Yau space $M$ by the map \bea \label{genm5map} \C-\{a_1, \ldots, a_k\}
\longrightarrow \Sigma \subset M, ~~~~~x \to (x, ~\sum_{k=1}^n\frac{\zeta_k}{x -a_k},
~~\prod_{k=1}^n(x-a_k)^{N_k}). \eea There are $(n+1)$ components at infinity and the bending behavior is as
follows: \bea
\label{m5prod} x\to \infty,~~y\to 0,~~~~t\sim x^{N_1 +\cdots + N_n}\\
\nonumber x\to a_k,~~y\to\infty,~~~~t\sim \zeta_k^{N_k} \prod_{i=1, i\neq k}^n (a_k - a_i)^{N_i} y^{-N_k}. \eea

We would like to identify the parameters $\zeta_k$ of (\ref{genm5map}) in terms of the scales of the ${\cal N} =
1$ field theory. We begin with the ${\cal N} = 2, U(N)$ curve which is \bea \label{n2} t^2 + B(x, u_k)) t +
\Lambda_{{\cal N}=2}^{2 N} = 0 \eea where we restored a dimensional dependence on the QCD scale $\Lambda_{{\cal N}=2}^{2 N}$.
We need to deform this curve to (\ref{genm5map}) which describes $\CN =1$ $U(N_1) \times \cdots \times U(N_n)$
supersymmetric gauge theory. We first consider a degeneration of (\ref{n2}) to the form \bea \label{prodn2} t^2
+t \prod_{k=1}^n (x- a_k)^{N_k} + \Lambda_{{\cal N}=2}^{2N} = 0. \eea  In the classical type IIA limit, this corresponds
to a brane configuration of two parallel NS branes and $n$ groups of $N_k$ coincident D4 branes located at
$x=a_1, \ldots a_n$ between two NS branes. There are $N$ such points in the Coulomb branch and each of them are
related by the discrete group $\Z_{2N_1} \times \cdots \times \Z_{2N_n}$ in which each factor acts on the QCD
scale as \bea \Lambda_{{\cal N}=2}^{2} \rightarrow e^{2 \pi i/N_k}
 \Lambda_{{\cal N}=2}^{2}. \eea

For $t\to \infty$, the degenerate  curve (\ref{prodn2}) is asymptotically $ t \sim x^N$ and to see the asymptotic
behavior for $t\to 0$ and $x\to a_k$, we rewrite (\ref{prodn2}) as \bea \label{prodn3} \frac{t}{\prod_{i=1, i\neq
k}^n (x- a_i)^{N_i}}+ (x- a_k)^{N_k} + \frac{\Lambda_{{\cal N}=2}^{2N}}{ t \prod_{i=1, i\neq k}^n (x- a_i)^{N_i}} = 0.
\eea As $t\to 0$ and $x\to a_k$, the degenerate curve (\ref{prodn3}) is asymptotically \bea x-a_k\sim
\frac{\Lambda_{{\cal N}=2}^{2N/N_k}}{ t^{1/N_k} \prod_{i=1, i\neq k}^n (a_k- a_i)^{N_i/N_k}} \eea At large $t$ we have a
curve approaching  $x = t^{1/N}$ and $y$ is small, while for $t \rightarrow 0$ the curve approaches $y_k = \mu_k
x$ for the $k$-th NS$'$ brane where $\mu_k$ is the mass of the $SU(N_k)$ adjoint. For the deformation (\ref{wt}),
the mass of the $SU(N_k)$ adjoint field is \bea \mu_k = g_{n+1} \prod_{i \ne k} (a_k - a_i) \eea so \bea
\label{y} y_k = g_{n+1} \Lambda_{{\cal N}=2}^{2 N/N_i} \prod_{i \ne k} (a_k - a_i)^{1 - N_i/N_k} t^{-1/N_k} \eea By
threshold matching between the high energy $SU(N)$ theory with the low energy theory obtained by integrating out
the adjoint fields and the massive W bosons, one obtains: \bea \label{scalem} \Lambda_k^{3} = g_{n+1}
\Lambda_{{\cal N}=2}^{2 N/N_i} \prod_{i \ne k} (a_k - a_i)^{1 - 2 N_i/N_k} \eea where $\Lambda_k$ is the scale of the
low-energy $SU(N_k)$ theory. From (\ref{y}) and (\ref{scalem}), we can rewrite $y_k$ as: \bea \label{yscale} y_k
= \Lambda_k^{3} \prod_{i \ne k} (a_k - a_i)^{N_i/N_k} t^{-1/N_k} \eea We can now compare the equations
(\ref{yscale}) and (\ref{m5prod}). The boundary condition for $x \rightarrow a_k$ is \bea \label{oyscale} y_k =
\zeta_k \prod_{i \ne k} (a_k - a_i)^{N_i/N_k} t^{-1/N_k} \eea We then see that $\zeta_k = \Lambda_k^{3}$ which,
as we would expect, looks the same as for the $SU(N)$ case.

If we now consider the fivebrane (\ref{genm5map}), we see that its asymptotics depend on $\zeta_i^{N_i}$ but the
brane itself depends on $\zeta_i$, so the M5 branes related by the $Z_{N_{1}} \times \cdots \times Z_{N_{n}}$
transformations: \bea \zeta_i \rightarrow e^{\frac{2 \pi i}{N_i}} \zeta_i,~~~~i=1,~\cdots,~n \eea have the same
asymptotic behavior but differ in  shape.
Therefore the $N_i$ possible value for each $\zeta_i$ corresponds to a
different M5 brane and hence  to a different vacuum of the quantum theory. We shall see below that these different
quantum states split after we go through a transition. The fact that the curve (\ref{genm5map}) has properties
which are just a simple generalization of the $SU(N)$ case makes it easier to give a description of confinement.

When the size of $\P^1$ goes to zero, the $x^7$ direction in the M-theory will be very small and negligible. The
modulus of $t$ on $\Sigma$ will be fixed i.e. $\Sigma$ will be a curve in the cylinder $\S^1 \times \C^2$ where
$\S^1$ is the circle in the 11-th dimension and $\C^2$ are coordinatized by $x, y$. In fact, the value of $t$ on
$\Sigma$ must be constant because
 $\Sigma$ is holomorphic and there is no non-constant holomorphic map
into $\S^1$. The M5 curve makes a transition
 from a space curve into a plane curve.
 We now need to eliminate $t$ since the dimension of $t$ on $\Sigma$ is
virtually zero and the information along $t$ is not reliable. In the process of eliminating $t$, there are $N_k$
possibilities of $\zeta_k$ since only $\zeta_k^{N_k}$, and not $\zeta_k$,
enters in the behavior at infinity. They
represent possible supersymmetric vacua in the gauge theory which is determined by the interior behavior of the
brane. Thus we obtain the following relation between $x$ and $y$: \bea yW'(x) = \sum_{k=1}^n \zeta_k \exp( 2\pi i
m_{k,i}/N_k) \prod_{l=1, l\neq k}(x-a_l), \eea where $m_k = 1, \ldots, N_k$ for $k =1,\ldots, n$. Since this is
the limit where $g_sN$ is big, this degenerate M5 brane should not be considered as a M theory lift of D branes.
In this limit, the metrically deformed background without the D-branes is the right description and this is a
closed string geometric background.

The T dual picture of the deformed conifold (\ref{cydef}) is exactly an M theory lift of the NS brane of the
deformed conifold with \bea \label{muk} \mu_k = \zeta_k \exp(2\pi i m_{k,i}/N_k) \prod_{l=1, l\neq k}(a_k-a_l).
\eea The size of $\S^3$ on the deformed conifold depends on the expectation value for the gluino condensation
and, for each value of the gluino condensate we will have different flux through the $\S^3$ cycle. We may
intuitively consider the plane M5 as one obtained from two intersecting M5 branes by smearing out the
intersection point due to the flux from the vanished D4 branes wrapped on  $\P^1$.

\subsection{Field theory analysis}
The $\CN =2$ pure $SU(N)$ Yang-Mills theory can be obtained by wrapping N D5 branes on the zero section of
$\CO(-2) +\CO(0)$ over $\P^1$.  After taking T-duality along the direction of the circle action $\CS_2$ of
(\ref{s2}) and lifting to M-theory, we obtain M5 curve of the form: \bea \label{Msunsw} t^2 + B(x)t + 1 =0\eea
where \bea B(x) = x^N + u_1 x^{N-1} + u_2 x^{N-2} + \ldots + u_k \eea (for $SU(N)$ we impose $u_1 =0$). The
deformation of the theory by the superpotential $\Wt$ has been achieved by perturbing the geometry from $\CO(-2)
+\CO(0)$ over $\P^1$ to a Calabi-Yau with $n$ $\P^1$'s with normal bundle $\CO(-1) + \CO(-1)$ which are located
at $x=a_1, \ldots, x_n$. In the T-dual type IIA picture, we have one NS branes and $n$ NS$'$ branes (Figure 2) and
the NS$'$ branes are located at $x=a_1, \ldots, x_k$. Hence classically we have \bea yW'(x) =1\eea where $y$ is the
coordinates of NS$'$ and $x$ is that of NS. The gauge group $U(N)$ is broken into $U(N_1) \times \cdots \times
U(N_n)$, and after gluino condensations, the final gauge theory is $\CN =1$ $U(1)^n$ theory. The $k$-th $U(1)$
gauge field of $U(1)^n$ theory  is given by the center of the mass coordinates of $N_k$ D4 branes along the NS
branes. Since the $N_k$ branes are coincident in the type IIA picture, the moduli of $U(1)^n$ is classically
described by the NS branes which is given by \bea \label{classu1nmod} yW'(x) =1. \eea But if we lift the branes
configuration to the M-theory, the quantum effects will show up and the corresponding M5 brane is given by an
embedding of the $n$-punctured $x$ plane into the Calabi-Yau space $M$ by the map \bea \label{genm5map2}
\C-\{a_1, \ldots, a_k\} \longrightarrow \Sigma \subset M, ~~~~~x \to (x, ~\sum_{k=1}^n\frac{\zeta_k}{x -a_k},
~~g_n\prod_{k=1}^n(x-a_k)^{N_k}). \eea Hence the MQCD moduli of the $U(1)^n$ is given by \bea \label{mqcdu1n}
yW'(x) = f_{n-1} (x, \zeta_1, \ldots , \zeta_n),
 \eea
which is the M-theory description of the NS branes. Here $f_{n-1}(x, \zeta_1, \ldots , \zeta_n)$ is given as in
(\ref{fn-1}).  As we have discussed in \cite{dot}, the effective low energy four dimensional theory comes from
the world volume of the M5 brane and the low effective action is determined by the Jacobian of the M5 brane.
Since (\ref{mqcdu1n}) is  a non-compact curve which is a $\P^1$ with $(n+1)$ points punctured, the Jacobian is an
algebraic group $(\C^{*})^*$ and \bea H^1((\C^{*})^n, \Z) = \oplus_{k=1}^n \C \frac{dx}{x-a_k}. \eea By putting
the cut-off at $|x| = \Lambda_{0,k}^{3/2}$, the integral over the $k$-th component of the Jacobian $(\C^{*})^n$
\bea \int_{\C^*} d \log x \wedge \ast d\log x \sim 3 \log \Lambda_{0,k} - \log |\zeta_k| \eea becomes finite and
the coupling constants of the $k$-th component of $U(1)^n$ is given by \bea \frac{1}{g_k^2} \sim 3 \log
\Lambda_{0,k} - \log |\zeta_k| \eea which is the expected running of the coupling constant if we replace the
cutoff $\Lambda_{0,k}$ by the scale of the gauge theory $\Lambda_k$.

On the other hand, the $\CN =2$  pure $SU(N)$ Yang-Mills theory deformed by a tree level superpotential
(\ref{wt}) only has unbroken supersymmetry on submanifolds of the Coulomb branch, where there are additional
massless fields besides the $u_k$. They are nothing but the magnetic monopoles or dyons which become massless on
some particular submanifolds $\ev{u_k}$ where the Seiberg--Witten curve degenerates. Near a point with $l$
massless monopoles, the superpotential is
\begin{equation}
W = \sqrt{2} \sum_{i=1}^{l} M_i A_i \tilde M_i + \sum_{k=1}^{n+1} g_{k} U_{k} ~,
\end{equation}
where $A_i$ are the chiral superfield parts of the $U(1)$ vector multiplet corresponding to an $\CN = 2$ dyon
hypermultiplet $M_i$ and $U_k$ are the chiral superfields representing the operators $\Tr (\Phi^k)$ in the low
energy theory. The vevs of the lowest components of $A_i,M_i,U_{k}$ are written as $a_i,m_{i},u_{k}$. The
supersymmetric vacua are at those $\ev{u_{k}}$ satisfying:
\begin{equation}
\label{vacua} a_i(\ev{u_{k}}) = 0 ~, ~~~~~~~ g_k + \sqrt{2} \sum_{i=1}^{l}\frac{\pa a_i}{\pa u_{k}}(\ev{u_{k}})
m_i \tilde m_i = 0 ~,
\end{equation}
for $k=1,\ldots,n+1$. The value of the superpotential at this vacuum is given by
\begin{equation}
\We=\sum _{k=1}^{n+1}g_k \ev{u_{k}} ~.
\end{equation}
Recall that the Seiberg-Witten curve of $\CN= 2$ $U(N)$ is \bea y^2 = P(x; u_r)^2 - 4 \Lambda^{2N}_{{\cal N} =2 }, P(x, u_r)
\equiv \det (x -\Phi) = \sum_{k=0}^N x^{N-k}s_k \eea where the $s_k$ are related to the $u_r$ by the Newton's
formula \bea ks_k + \sum_{r=1}^{k} ru_r s_{k-r} =0, \eea and $s_0\equiv 1$ and $u_0 \equiv 0$.  The condition for
having $N-n$ mutually local massless magnetic monopoles is that \bea \label{swfac} P_N(x, \ev{u_p})^2  - 4
\Lambda^{2N}_{{\cal N} =2} = (H_{N-n}(x))^2 F_{2n}(x) \eea where $H_{N-n}$ is a polynomial in $x$ of degree $N-n$ with distinct
roots and $F_{2n}$ is a polynomial in $x$ of degree $2n$ with distinct roots.

On the $n$ massless photons, the one corresponding to the trace of $U(N)$, does not couple to the rest of the
theory and so its coupling constant is the one we started with. The other $n-1$ photons which are left massless
after the breaking $U(N) \to U(N_1)\times \cdots \times U(N_n)$ have gauge couplings which are given by the
period matrix of the reduced curve \bea \label{swu1n} y^2 = F_{2n}(x; \ev{u_r}) = F_{2n} (x; g_k,\Lambda)\eea
with $F_{2n}(x; \ev{u_k})$ the same functions appearing in (\ref{swfac}) and $\ev{u_r}$ the point of the solution
space of (\ref{swfac}) which minimize $\Wt$. The curve (\ref{swu1n}) thus gives the exact gauge couplings of
$U(1)^{n-1}$ which remains massless in (\ref{vacua}) as function of $g_k$ and $\Lambda$.

Comparing the results from the MQCD, the curve (\ref{swu1n}) cannot be isomorphic to the curve (\ref{mqcdu1n})
because, for $n>1$, the curve (\ref{swu1n}) is hyperelliptic  while the curve (\ref{mqcdu1n}) is rational. But
two curves are intimately related. We may rewrite (\ref{swu1n}) as \bea y^2 = W'(x)\left( W'(x) +
\frac{f_{n-1}(x)}{W'(x)}\right). \eea From this form, it is clear that one can obtain (\ref{mqcdu1n}) from
(\ref{swu1n}) and vice versa in a canonical way. Thus the moduli (\ref{swu1n}) and (\ref{mqcdu1n}) are
equivalent. This also can be traced back to the fact that our model is based on the singular conifold whose equation
is given by \bea W'(x)y -uv =0. \eea If we now replace the above equation by \bea W'(x)^2 +y^2 -uv =0, \eea we obtain
the same model and after the transition the geometric background for the closed string will be given by \bea
W'(x)^2 +f_{n-1} +y^2 -uv = 0 \eea which is the original form given in \cite{civ}. Via T-duality, the geometric
background is given as an NS brane wrapping on \bea W'(x)^2+ f_{n-1}(x) + y^2 =0.\eea Thus the moduli will be
described by \bea \label{geomu1n}W'(x)^2 +f_{n-1}(x) +y^2 = 0,\eea and we can prove two curves (\ref{swu1n}) and
(\ref{geomu1n}) are isomorphic i.e. \bea W'(x)^2 + f_{n-1}(x) = -g^2_{n+1} F_{2n}(x) \eea after possible
coordinates changes. This was proved up to a certain order of $x$ in \cite{civ}. But we do not expect to have a
proof via M-theory because the M5 curve for $\CN =1$ is always rational. It is a hyperelliptic curve only in the
case of $\CN =2$. The moduli (\ref{swu1n}) looks like a $\CN=2$ curve in a sense that the curve is a plane curve
rather than a space curve which is also reflected by the fact that we have massive glueballs.

\section{Adding Matter Fields and Orientifold Planes}
We can add some quark chiral superfields in the fundamental representation of $U(N_k)$ for $k=1, \ldots , n$
generalizing the discussion of \cite{civ}.  In the type IIB, the matter fields are added as D5 branes wrapping a
holomorphic 2-cycle separated by a distance $m$ from the exceptional $\P^1$. There are two kinds of matter fields
we may add. In the T-dual type IIA picture, the matter fields are obtained by adding semi-infinite D4 branes
which can be attached either the NS brane or the NS$'$ branes in the Figure 2.  Recall that the small resolution
$\tilde{\CC}$ is covered by two copies of $\C^3$,  $U_p, p=0,1$ with coordiantes $Z, X, Y$ (resp. $Z', X' , Y'$).
The semi-infinite D4 branes attached to the NS brane near the vacua $x = a_k$ is given by the D5 branes wrapping
non-compact holomorphic cycles  \bea Y = 0,~~X - a_k =m_{k,1}, \ldots , m_{k, F_k}. \label{nfcycle1}, \eea
We can
also put
 the semi-infinite D4 branes near $x= a_k$
attached to the NS$'$ brane  which are given by the D5 branes wrapping a non-compact holomorphic 2-cycles \bea Y' =
0,~~X' -a_k = n_{k,1}, \ldots , n_{k, G_k}. \label{nfcycle2}\eea

These semi-infinite D4 branes give rise to $N_{k,f}$ hypermultiplets in the fundamental representation of
$U(N_k)$ corresponding to strings stretched between the $N_k$ D5 branes wrapping the exceptional $\P^1$ at $x
=a_k$ and the $N_{k,f}$ non-compact D5 branes wrapping the holomorphic 2 cycles given by (\ref{nfcycle1}) and
(\ref{nfcycle2}) and the mass is the distance between the non-compact D5 brane and the compact D5 wrapping
$\P^1$. Thus we can geometrically engineer the $\CN =1$ $U(N_1) \times \cdots \times U(N_n)$ theory with adjoints
and $N_{k,f}$ hypermultiplets with mass $m_{k,1} , \ldots ,m_{k, N_{k,f}}$.

We remark that the distance between non-compact holomorphic 2-cycles defined (\ref{nfcycle1}) (resp.
(\ref{nfcycle2})) asymptotically approach to a fixed cycle.  To see the asymptotic behavior of the cycles defined
by (\ref{nfcycle1}), consider its defining equations in the open set $U_1$: \bea X' -a_k = m_{k,1}, \ldots ,
m_{k, F_k}, Y' = - W'(a_k + m_{k,1})Z'^{-1}, \ldots ,- W'(a_k + m_{k, F_k})Z'^{-1} .\eea Hence as $\Z \to \infty$
all cycles approaches to a cycle given by \bea X' -a_k = m_{k,1}, \ldots , m_{k, F_k},~~ Y' =0,\eea and thus
there is no transversal oscillation at infinity. In the type IIA picture, they begin from NS brane at one side
and tangentially approach to the other NS brane. So beyond a cut-off $|Z| = \Lambda_0$, the oscillations in the
transversal direction can be ignored and the strings between them see only 4 infinite directions.  This is
important when we discuss the Seiberg duality and we impose the condition that in the magnetic picture the mesons
live in a four dimensional theory. The same aspect was discussed in \cite{SS} where additional NS branes were
needed in order to make the flavor D4 branes very long but finite. In \cite{mv1} a similar consideration has been
utilized in calculating the contribution of massive flavor to the superpotential.

We can also discuss the $SO/Sp$ theories by introducing additional orientifolds in the theory. The orientifolding
can be introduced by extending the complex conjugation on the conifold defined by \bea z_1^2 + z_2^2 +z_3^2 +
z_4^2 =0 \eea to both complex and K\"ahler deformations. On the complex deformed conifold $T^*\S^3$, the special
Lagrangian $\S^3$ is invariant under this complex conjugation and hence the orientifold is an O6 plane wrapping
$\S^3$ \cite{sv,gomis}. In MQCD, the theory with the superpotential can be obtained by rotating the NS branes as
in \cite{aot1}.

We discuss now the type IIB picture by first reviewing the results of \cite{eot}. We shall concentrate on the
case ${\cal N} = 1, SO(N)$ gauge theory with matter $\Phi$ in the adjoint representation of $SO(N)$ and with the
superpotential \bea \label{sotp} W_{SO,tree}(\Phi) = \sum_{k=1}^{n+1} \frac{g_{2 k}}{2 k} Tr \Phi^{2 k} \eea The
classical theory with the superpotential (\ref{sotp}) has vacua where the eigenvalues of $\Phi$ are roots of \bea
W'(x) = g_{n+1} x \prod_{j=1}^{n} (x^2 + a_j^2),~~~a_j~>~0 \eea In \cite{eot}, we considered D5 on $\P^1$ cycles
located at the zeros of $W'(x)$. In the presence of an orbifold plane located at $x=0$, the $\P^1$ cycle at $x =
0$ becomes an $\RP^2$ cycle stuck on the orientifold plane and the field theory on the D5 branes wrapped on it is
$O(2 N_0)$. The D5 branes on $\P^1$ cycles located at $i~a_j$ are identified with the D5 branes on $\P^1$ cycles
located at $-i~a_j$ and the field theory on the corresponding D5 branes is $U(N_j)$, so the breaking of the gauge
group is \bea O(N) \rightarrow O(N_0) \times U(N_1) \times \cdots \times U(N_n) \eea If we blow down the
geometry, it becomes a Calabi-Yau with conifold singularities (as in $SU(N)$ case), and we consider a small
resolution of it as in (\ref{cy}). We take the action of the circle action $\CS_{c}$ on $\tilde{\CC}_{SO}$, where
$\tilde{\CC}_{SO}$ is the small resolution for the $SO$ case and the same arguments of section 3.1 tell us that
the T-dual picture is a brane configuration with one $NS_x$ brane and $2 (n +1)$ $NS_{Y'_{i}}, i=0,\cdots,2~n~+1$
located  at $x = 0, x = \pm i a_j, j = 1, \cdots n$, $N_i, i=1,~\cdots,~2~n~+~1$ D4 branes connecting the $NS_x$
brane with an $NS_{Y'_{i}}$ brane and $N_0$ D4 branes connecting the $NS_x$ brane with the  $NS_{Y'_{0}}$ brane.
Under the orientifold action, the pairs of $NS_{Y'_{i}}$ branes located at $\pm i a_j, j = 1, \cdots n$ are
identified and give the $ U(N_1) \times \cdots \times U(N_n)$ group and the D4 branes which connect the $NS_x$
with $NS_{Y'_{0}}$ brane and stay on top of the orientifold plane give $O(N_0)$ group.

We can also lift the configuration to M theory. The complex curve $\Sigma$ is the embedding of the complex plane
coordinatized by $x$ and punctured at $0,~\pm i a_i, i=1,\cdots,N$ into the Calabi-Yau space M by the map: \bea \C
- \{0,\pm i a_1,\cdots,\pm i a_n\} \rightarrow \Sigma \subset M,~~~~~~ x \rightarrow\left(x, \frac{\zeta_0}{x} +
\sum_{k=1}^{n} \frac{\zeta_k}{x^2 + a_i^2}, \prod_{k=1}^{n} (x^2 + a_k^2)^{N_k}\right) \eea We could again
identify the parameters $\zeta$ of the above curve with parameters of field theory by studying a degeneration of
the Seiberg-Witten curve.

By considering now the limit when the size of the $\P^1$ cycle goes to zero, the M5 brane curve makes the
transition from a space curve into a plane curve. which is the M theory lift of the NS brane of the orbifolded
deformed conifold: \bea \mu_k = \zeta_k \exp(2\pi i m_{k,i}/N_k) \prod_{l=1, l\neq k}(a_k^2~-~a_l^2) \eea for the
$\S^3$ cycles which do not sit at the origin and \bea \mu_0 = \zeta_0 \exp(2\pi i m_{0,i}/(N_0/2 \pm 1))
\prod_{l=1} a_l^2 \eea for the $S^3$ cycle which sits at the origin. The values of $mu$ parameters appear in the
plane M5 brane defining equations and are connected to the fluxes over the $\S^3$ cycles in type IIB picture.

The field theory analysis goes as in section 3.3.

\section{Seiberg Duality versus Abelian Duality}
\subsection{$SU(N)$ group with $F$ flavors}
We discuss now the well known Seiberg duality \cite{seid} in our framework. This states that the infrared
behavior of $SU(N)$ theories with $F$ flavors of quarks $Q_+$ and antiquarks $Q_-$ have a dual description in
terms of an  $SU(\tilde{N})$ gauge group,  $\tilde{N} = F - N$, with F flavors of quarks $\tilde{Q_+}$ and
antiquarks $\tilde{Q_-}$, with gauge-singlet mesons $M \equiv Q_+Q_-$. The dual effective theory for massive
quarks $Q$ (with equal mass $m$, for simplicity) has a superpotential \bea \label{seisuper} W_{\mbox{dual}} =
\mbox{tr}~ mM + \frac{1}{\mu} \tilde{Q_+}M~\tilde{Q_-} \eea and the strong interaction scale of the dual theory
is \bea \label{dualscale} \tilde{\Lambda}^{3 \tilde{N} - F} \Lambda^{3 N - F} = (-1)^{\tilde{N}} \mu^{F} \eea
where the scale $\mu$ appears in order to compensate the difference in dimension of the scales, and also the
meson fields between the original and dual theories.

We will show that the Seiberg duality can be achieved by a birational flop\footnote{In algebraic geometry, this
type of geometric transition is usually called  flop. But recently in high energy physics, many other types of
geometric transitions have been called  flop. To distinguish, we call it  birational flop.} in the the geometric
engineering. To make it more clear, we first recall the flop process~\cite{deot}. The conifold \bea \CC:~~xy -uv
=0 \eea have two small resolutions which will be denoted by $\CC_x$ and $\CC_y$:

{$\bullet$ The small resolution $\CC_x$}

Let $\C^3_0$ and $\C^3_\infty$ be two $\C^3$ whose coordinates are $Z_1, X_1, Y_1$ and $Z'_1, X'_1, Y'_1$ with
identification: \bea Z'_1 =1/Z_1,~~X'_1 = X_1Z_1,~~Y'_1 = Y_1Z_1. \eea The blowing down map is given by \bea
\sigma_x:&&\CC_x = \C^3_0 \cup \C^3_\infty \longrightarrow \CC \\
\nonumber &&x= X_1,~~ y= Y_1Z_1,~~u=Y_1,~~v=X_1Z_1\\ \nonumber &&x= X'_1Z'_1,~~y=Y'_1,~~u=Y'_1Z'_1,~~v= X'_1.\eea

{$\bullet$ The small resolution $\CC_y$}

Let $\C^3_0$ and $\C^3_\infty$ be two $\C^3$ whose coordinates are $Z_2, X_2, Y_2$ and $Z'_2, X'_2, Y'_2$ with
identification: \bea Z'_2 =1/Z_2,~~X'_2 = X_2Z_2,~~Y'_2 = Y_2Z_2. \eea The blowing down map is given by \bea
\sigma_y:&& \CC_y = \C^3_0 \cup \C^3_\infty \longrightarrow \CC \\ \nonumber &&x= X_2Z_2,~~y=Y_2,~~u=X_2,~~v=
Y_2Z_2\\
\nonumber &&x= X'_2,~~ y= Y'_2Z'_2,~~u=X'_2Z'_2,~~v=Y'_2 .\eea

The exchange of the role of $x$ and $y$ in the resolution induces an isomorphism $f: \CC_x \to \CC_y$ given by:
\bea \label{flop} Z_1 \mapsto Z_2,~X_1 \mapsto Y_2~Y_1 \mapsto X_2,~Z'_1\mapsto Z'_2,~X'_1\mapsto
Y'_2,~Y'_1\mapsto X'_2. \eea We may define the circle action on $u, v$ variables as before. In the T-dual picture
along the direction of the orbits of the circle action, the flop will take the $N_x$ (resp. $N_y$) brane (which
is spaced along the $x$ (resp $y$)-direction) of $\CC_x$ to the $N_y$ (resp. $N_x$) brane (which is spaced along
the $y$ (resp. $x$)-direction) of $\CC_y$. Nevertheless, it seems to be impossible to argue the Seiberg duality
in the geometric engineering or in the brane set-up without going to the M-theory.

In the previous section we have discussed the realization in geometry for $SU(N)$ with fundamental matter. This
is realized by having $N$ D5 branes wrapped on the exceptional $\P^1$ cycles and $F$ D5 branes wrapped on the
2-cycle (\ref{nfcycle1}),which gives by T-duality a brane configuration with two `orthogonal'  NS branes,
together with $N$ D4 branes between them on an interval and $F$ semi-infinite D4 branes attached to one of the NS
branes. The positions of the semi-infinite D4 branes on the NS branes determine the bare masses of the quark
fields. This classical string theory brane configuration can be lifted to M5 curve $\Sigma$  as an embedding of
the punctured $x$ plane into the Calabi-Yau space $M$: \bea \label{melectric} x ~\to ~\left(x,~ \zeta x^{-1},
~\xi\frac{x^N}{\prod_{i=1}^{F} (x - m_i)} \right) \eea where the parameters $\xi$ and $\zeta$ are identified with
\bea \label{parelectric} \xi = (\prod_{i} m_i)^{(F - N)/F}~,~~\zeta = \Lambda^{(3 N - F)/N} (\prod_{i}
m_i)^{1/N}.\eea

Now the birational flop will change the role of $x$ and $y$, and hence the same M5 curve $\Sigma$ can be
considered as an embedding of the punctured $y$ plane into the Calabi-Yau $M$~\cite{bisky,SS}: \bea
\label{mmagnetic} y ~\to ~\left(\tilde{\zeta} y^{-1},~ y, ~\tilde{\xi}(-1)^{F}
\frac{y^{\tilde{N}}}{\prod_{i=1}^{F} (y - \ev{M}_i))} \right) \eea where the eigenvalues of the meson fields are
given by \bea \ev{M}_i = \frac{(\prod_{i} m_i)^{1/N} \Lambda^{(3 N - F)/N}}{m_i}, \eea and the new parameters are
where \bea \label{parmagnetic} \tilde{\xi}~=~(\prod_i \ev{M}_i)^{{F - \tilde{N}}/F},~~ \tilde{\zeta}~=~(\prod_i
\ev{M}_i)^{1/{\tilde{N}}} \tilde{\Lambda}^{(3~\tilde{N}-F)/\tilde{N}}. \eea Here the new set of parameters are
introduced in order to have field theory interpretation. Going from (\ref{melectric}) to (\ref{mmagnetic}) means
to  go between the two sides of the Seiberg duality. Therefore, the birational flop induces the Seiberg
duality. This M theory transition has been noticed in \cite{bisky, SS}.

We can now discuss of what happens in the limit of very small size of the $\P^1$ cycles where the $N$ D5 branes
corresponding to color gauge group are wrapped. In this limit the value of t on $\Sigma$ is again a constant and
both curved M5 branes (\ref{melectric}),(\ref{mmagnetic}) have a transition to plane M5 branes as discussed in
\cite{dot}. For the electric theory, in order to obtain the description of the plane M5 curves in this case, we
first fix $t$ at $t_0$ and the equation (\ref{melectric}) becomes: \bea t_0 = \xi \frac{x^N}{\prod_{i=1}^{F} (x -
m_i)}, t_0^{-1} = \zeta^{-N} \xi^{-1} y^N \prod_{i=1}^{F} (x - m_i) \eea which tells us that for the electric
case there are $N$ possible plane curves that the M5 space curve $\Sigma$ can be reduced to: \bea \label{planeel}
t~=~t_0,~~~x~y~=~\zeta~e^{2 \pi i k/N},~~ k = 0,1,\cdots,N-1 \eea They correspond to the $N$ vacua obtained after
integrating our the massive flavors and gluino condensation as: \bea \label{sgelectric} \ev{S} = exp(2 \pi i k/N)
\zeta,~~ \mbox{for electric theory} \eea

The same thing is true for the magnetic theory, where we fix $t = t_0$ and the equation (\ref{mmagnetic}) becomes
\bea t_0 = \tilde{\xi}(-1)^{F} \frac{y^{\tilde{N}}}{\prod_{i=1}^{F} (y - \ev{M}_i)}, t_{0}^{-1} = \tilde{\xi}
(-1)^{-F} \tilde{x}^{\tilde{N}} \prod_{i=1}^{F} (y - \ev{M}_i) \eea which tells us that for the magnetic case
there are $\tilde{N}$ possible plane curves that the M5 space curve can be reduced to: \bea \label{planemag}
t~=~t_0,~~~\tilde{x}~\tilde{y}~=~\tilde{\zeta}~e^{2 \pi i k/\tilde{N}},~~ k = 0,1,\cdots, \tilde{N}-1 \eea They
correspond to the $\tilde{N}$ vacua obtained after integrating our the massive mesons and gluino condensation as:
\bea \label{sgmagnetic} \ev{S} = exp(2 \pi i k/\tilde{N}) \tilde{\zeta},~~ \mbox{for magnetic theory} \eea From
equations (\ref{parmagnetic}) and (\ref{parelectric}) we know how to express $\zeta$'s in terms of the scales
$\Lambda$ so the equations for $\ev{S}$ agree with the field theory results: \bea \ev{S}^N =  \Lambda^{(3 N - F)}
(\prod_{i} m_i),~\mbox{for electric theory} \label{selectric} \eea and \bea \label{smagnetic} \ev{S}^{\tilde{N}}
= ~\tilde{\Lambda}^{3~\tilde{N}-F} (\prod_i \ev{M}_i),~ \mbox{for magnetic theory} \eea In \cite{dot} we have
discussed the reduction of the plane M5 branes to a configuration T-dual to the deformed conifold with fluxes.
The flux through the $\S^3$ cycle was related to the parameters appearing in the right hand side of
(\ref{sgelectric}) and (\ref{sgmagnetic}).

In other words, the fluxes through the $\S^3$ cycle would tell us whether we are in the electric theory or in the
magnetic theory. Therefore the geometry not only encodes the information regarding the gluino condensation in the
field theory but also captures
 the information regarding the Seiberg duality for the case of massive
matter! We expect this to be a a feature of a more general identification of field theory strong coupling results
in the geometry.

What about the $U(1)$ groups on the plane M5 branes? In \cite{dot}, we used a cutoff on the integral
$\int_{\Sigma} d~log~x \wedge *~d~log~x$ which was then identified with the scale of the gauge theory. In
discussing the Seiberg duality, there is one energy scale for the electric theory $\Lambda$ and one energy scale
for the magnetic theory $\tilde{\Lambda}$ which are related by (\ref{dualscale}). This means that we replace the
integral cutoff by $\Lambda$ for the electric theory and by $\tilde{\Lambda}$ for the magnetic theory. Because we
do not have any particle charged under the $U(1)$ groups, this identification is a sign of an abelian
electric-magnetic duality for the $U(1)$ groups. Therefore, after the transition, the nonabelian
electric-magnetic Seiberg duality gives rise to an abelian electric-magnetic duality. It should be very
interesting to study models where there are particles charged under the remaining $U(1)$ groups and to study the
corresponding match under the abelian electric-magnetic duality.
\subsection{$SO/Sp$ Groups}
The Seiberg duality has been interpreted in M theory for $SO/Sp$ groups in \cite{csaski} as an interpolation
between $m_i$ and $\ev{M}_i$. The flop transition discussed in this section is valid for the orbifolded conifold
too and the duality can be seen as such a flop. The Seiberg duality for $SO(N)$ groups states that an electric
$SO(N)$ with $F$ flavors in the vector representation is dual to a magnetic description with gauge group $SO(F -
N - 4)$ and the one for $Sp(2 N)$ states that an electric $Sp(2 N)$ with $2 F$ flavors in the fundamental
representation is dual to a magnetic description with the gauge group $Sp(2~F~-~2~N~-~4)$.

In terms of embedding of the punctured x plane into the Calabi-Yau space M, the M5 brane curves describing the
electric theory are

$\bullet$ for $Sp(2 N)$ groups with $2 F$ fundamental fields, the electric description is \bea
\label{melectricsp} x ~\to ~\left(x,~ \zeta x^{-1}, ~\xi\frac{x^{2 N +2}}{\prod_{i=1}^{F} (x^2 - m_i^2)} \right)
\eea where the parameters $\xi$ and $\zeta$ are identified with \bea \label{parelectricsp} \xi =
(\mbox{Pf}~m)^{2(F - N- 1)/F}~,~~\zeta = \Lambda^{(3(N+1) - F)/(N+1)} (\mbox{Pf}~m )^{\frac{1}{N+1}}.\eea and the
magnetic description is \bea \label{mmagneticsp} y ~\to ~\left(\tilde{\zeta} y,~ y, ~\tilde{\xi} \frac{y^{2
\tilde{N} + 2}}{\prod_{i=1}^{F} (y^2 - \ev{M}_i^2))} \right) \eea where the eigenvalues of the meson fields are
given by \bea \ev{M}_{ij} = [2^{N-1} (\mbox{Pf}~m)^{1/(N+1)}] \Lambda^{(3 (N+1) - F)/(N+1)}(\frac{1}{m})_{ij},
\eea and the new parameters are \bea \label{parmagneticsp} \tilde{\xi}~=~(\mbox{Pf} \ev{M})^{\frac{2(F -
\tilde{N} - 1)}{F}},~~ \tilde{\zeta}~=~(\mbox{Pf} \ev{M})^\frac{1}{\tilde{N}+1}
\tilde{\Lambda}^{(3(\tilde{N}+1)-F)/(\tilde{N}+1)}. \eea

$\bullet$ for $SO(2 N)$ groups with $2 F$ vectors, the electric description is \bea \label{melectricso} x ~\to
~\left(x,~ \zeta x^{-1}, ~\xi\frac{x^{2 N - 2}}{\prod_{i=1}^{F} (x^2 - m_i^2)} \right) \eea where the parameters
$\xi$ and $\zeta$ are identified with \bea \label{parelectricso} \xi = (\mbox{det} m)^{(F - N + 1)/F}~,~~\zeta =
\Lambda^{(3(N - 1) - F)/(N-1)} (\mbox{det} m )^{\frac{1}{2 N - 2}}.\eea and the magnetic description is \bea
\label{mmagneticso} y ~\to ~\left(\tilde{\zeta} y,~ y, ~\tilde{\xi} \frac{y^{2 \tilde{N} - 2}}{\prod_{i=1}^{F}
(y^2 - \ev{M}_i^2))} \right) \eea where the eigenvalues of the meson fields are given by \bea \ev{M}_{ij} = [16
(\mbox{det}~m)^{1/(2 N - 2)}] \Lambda^{(3 (N - 1) - F)/(N - 1)}(\frac{1}{m})_{ij}, \eea and the new parameters
are \bea \label{parmagneticso} \tilde{\xi}~=~(\mbox{det} \ev{M})^{\frac{(F - \tilde{N} + 1)}{F}},~~
\tilde{\zeta}~=~(\mbox{det} \ev{M})^\frac{1}{2 \tilde{N} - 2}
\tilde{\Lambda}^{(3~(\tilde{N}-1)-F)/(\tilde{N}-1)}. \eea

All the M5 branes (\ref{melectricsp}),(\ref{mmagneticsp}), (\ref{melectricso}),(\ref{mmagneticso}) have
transitions to plane M5 branes. For $Sp(2 N)$ groups, after choosing a constant value for $t~=~t_0$, $\Sigma$ for
the electric case becomes: \bea \label{plelsp} t~=~t_0,~~~x~y~=~\zeta~e^{2 \pi i k/(2 N + 2)},~~ k =
0,1,\cdots,2~N~+~1 \eea whereas for the magnetic case it looks like \bea \label{plmagsp}
t~=~t_0,~~~\tilde{x}~{y}~=~\tilde{\zeta}~e^{2 \pi i k/(2~\tilde{N}~+~2)}, ~~ k = 0,1,\cdots, 2~\tilde{N}~+~1 \eea
For the $SO(2~N)$ groups, after choosing a constant value for $t~=~t_0$, $\Sigma$ for the electric case becomes:
\bea \label{plelso} t~=~t_0,~~~x~y~=~\zeta~e^{2 \pi i k/(2~N~-~2) },~~ k = 0,1,\cdots,2~N~-~3 \eea whereas for
the magnetic case it looks like \bea \label{plmagso} t~=~t_0,~~~\tilde{x}~\tilde{y}~=~\tilde{\zeta}~e^{2 \pi i
k/(2~\tilde{N}~-~2)},~~ k = 0,1,\cdots, 2~\tilde{N}~-~3 \eea We see that for the $Sp(2 N)$ theories we have $2 N
+ 2$ vacua and for the $SO(2~N)$ theories we have $2 N - 2$ vacua. We encounter here the same puzzle which arises
in \cite{sv,widec} in what concerns the number of vacua of the field theory which can be read from the string
theory arguments. In \cite{sv}, the results depended on the normalization of the superpotential for the gluino
condensate.

We can also discuss the abelian electric-magnetic duality as we did for the $SU$ groups. On the worlvolume of the
plane M5 branes live $U(1)$ groups which are broken to ${\bf Z_2}$ groups by the orientifold projections. The
cut-offs are again related to the scales for the electric and magnetic theories, this is being a sign of the
abelian electric-magnetic duality obtained after the transition.

\subsection{Product of $SU(N)$ groups}
We now consider the discussion for product of $SU(N)$ groups with matter in the fundamental representation. We
use results of \cite{intri,gp,nos,sin}. The result is that if we start with an electric description of a field
theory with the gauge group $SU(N) \times SU(N')$ with $F (F')$ fundamental and $F (F')$ antifundamental flavors
with respect to the two gauge groups, together with bifundamental fields $Y$ with a superpotential $W~=~\mbox{Tr}
(X~\tilde{X}^2$, there is a magnetic description which has a dual gauge group $SU(2~F'+F- N') \times SU(2~F+F'-
N)$ with singlet fields that are a one-to-one map of the mesons of the electric theory. The dual theory has $F
(F')$ fundamental and $F (F')$ antifundamental flavors with respect to the two gauge groups, two bifundamentals
and several types of gauge singlet mesons.

The brane configuration corresponding to the field theories were first discussed in \cite{bh} with matter given
by D6 branes, in \cite{gp,sin} a configuration with semiinfinite D4 branes was used instead. In our discussion we
use a different brane configuration than \cite{gp,sin}, with the finite D4 branes ending on left on the same NS
brane. This brane configuration is lifted to M theory to an M5 brane given by (\ref{genm5map}) and (\ref{m5prod})
for $n=2$ and $F = F' = 0$.

It is possible to identify the parameters of the M5 branes in terms of the ${\cal N} = 1$ scales of field theory
by the same methods used for the case without matter. By doing that, one could again describe the field theory
vacua in terms of plane M5 branes after the geometrical transition.

\section{Acknowledgment}
We would like to thank Aki Hashimoto, Gary Horowitz, Ken Intriligator, Juan Maldacena, Carlos Nunez and Cumrun
Vafa for helpful discussions. The research of KD is supported by Department of Energy Grant no.
DE-FG02-90ER40542. The research of KO is supported by NSF grant no. PHY 9970664. The research of RT is supported
by DFG.

\end{document}